# THE IMPRINT OF Ω ON THE COSMIC MICROWAVE BACKGROUND

Naoshi Sugiyama[1,2] and Joseph Silk[1]

[1] *Departments of Astronomy and Physics, and Center for Particle Astrophysics, University of California, Berkeley, CA 94720*

[2] *Department of Physics, Faculty of Science, University of Tokyo, Tokyo 113, Japan*

## ABSTRACT

We investigate the dependence of large angular scale cosmic microwave background anisotropies on various initial conditions, including both adiabatic and isocurvature perturbations and the initial power law index $n$, in a variety of low-Ω cosmological models. Cosmological constant–dominated flat models and inflationary open adiabatic models, with $n_{\rm eff} < 1$, and open isocurvature models, with $n_{\rm eff} \simeq 2$, are significantly constrained by the present observations.

e-mail: sugiyama@pac3.berkeley.edu
silk@pac2.berkeley.edu

# I. Introduction

The discovery of cosmic microwave background (CMB) anisotropies by the COBE satellite [1] has provided important information about initial conditions in terms of the spectrum of primordial density perturbations. In particular, these observations provide a powerful probe of the slope of the initial power spectrum. The first year DMR data suggests an initial power law index $n = 1.1 \pm 0.5$, also in agreement with one recent analysis of second year combined data ($n = 1.10 \pm 0.32$) [2], that is consistent with the scale–invariant prediction $n = 1$ of inflationary cosmology. An independent analysis of two year data [3], however, may result in a steepening to $n = 1.59^{+0.49}_{-0.55}$, the recent detection of CMB anisotropies by the Tenerife experiment [4] supporting the higher power law slope. A value of $n$ significantly greater than unity is difficult to reconcile with inflation. In this paper, we consider a set of alternative models that naturally allow $n > 1$.

Previous discussions of large angular scale CMB anisotropies almost invariably adopt the flat $\Omega = 1$ universe with initially adiabatic perturbations. In an open universe, the shape of the CMB anisotropies induced by adiabatic perturbations is affected by background curvature on large scales [5,6,7]. It is generically difficult to disentagle curvature effects from initial conditions. However we have found that primordial isocurvature perturbations in an open, baryon–dominated (BDM) universe [8] result in a spectral shape that is primarily determined by background curvature. The interpretation of smaller angular scale observations are confused by the sub–horizon microphysics of Doppler peaks and rescattering; hence we focus here on large angular scales ($\gtrsim 5$ degrees). In this *Letter*, we present the results of numerical calculations of large-scale CMB anisotropies for both adiabatic and isocurvature initial conditions, in order to see whether the effects of geometry can be distinguished from the dependence on fluctuation mode, $\Omega$, initial power law index $n$, and thermal history of the universe.

## II. Large–Scale CMB Anisotropies

Here for generality, we consider open universe models. The temperature anisotropy is usually expanded into multipole components $\ell$ in Fourier space. Detailed treatments are given elsewhere [5,9]. In order to directly compare the spectrum with specific observations, we introduce the coefficients $C_\ell$[10] of the CMB anisotropies in $\ell$ space as $C_\ell/4\pi \equiv <|a_\ell(\eta_0)|^2>$, where $a_\ell(\eta_0)$ is the coefficient of the $\ell$th multipole component of the temperature anisotropy at present and $\eta_0$ is the present conformal time. The expected temperature anisotropy for each experiment is expressed by using $C_\ell$ and the specific window function $W_\ell$: $(\delta T/T)^2_{exp} = \sum_{\ell \geq 2} W_\ell (2\ell+1) C_\ell/4\pi$.

There are several physical contributions to CMB anisotropies for generic density perturbations [11], i.e., the Sachs-Wolfe (SW) effect, the integrated Sachs-Wolfe (ISW) effect, primordial entropy perturbations, induced Doppler effect and primordial adiabatic perturbations. On very large scales, the Doppler and adiabatic terms can be neglected. Hence the $\ell$th moment of the temperature anisotropy is written in gauge–invariant variables [12] as

$$a_\ell(\eta_0) = \left(\Psi(\eta_{LSS}) - \frac{1}{k}\frac{a'}{a}V(\eta_{LSS})\right) X^\ell_\nu(\eta_0 - \eta_{LSS})$$
$$+ 2\int_{\eta_{LSS}}^{\eta_0} \left(\frac{d\Psi}{d\eta}\right) X^\ell_\nu(\eta_0 - \eta) d\eta - \frac{S}{3} X^\ell_\nu(\eta_0 - \eta_{LSS}),$$

where $\Psi$, $V$, $S$ and $\eta_{LSS}$ are the gravitational potential, the velocity perturbation, the entropy perturbation and conformal time on the last scattering surface. $X^\ell_\nu(\eta_0 - \eta)$ is a radial eigenfunction in the open universe with $\nu = \tilde{k}/\sqrt{-K}$, where $K$ is the curvature constant and $\tilde{k} \equiv \sqrt{k^2 + K}$, with $k$ being wave number. If flat universe models are assumed, $X^\ell_\nu$ reduces to the usual spherical Bessel function $j_\ell(k(\eta_0 - \eta))$. Each component in this equation corresponds to SW, ISW and entropy terms, respectively. In the case of flat $\Omega = 1$ models, the SW term is $\Psi/3$ and the ISW term vanishes for growing modes in the matter–dominated universe. For adiabatic initial conditions, the entropy term is negligible. On the other hand, for isocurvature initial conditions, the combination of entropy term and SW terms becomes $2\Psi$ [13]. For $\Omega = 1$ models, one can easily calculate large–scale CMB fluctuations from the matter power spectrum through the Poisson equation. The relation between $C_\ell$ and the initial power law index $n$ of the matter spectrum is $C_\ell \propto \Gamma(\ell + (n-1)/2)/\Gamma(\ell - (n-5)/2)$; ($n < 3$) [6]. For isocurvature models, $n$ is replaced by $n+4$ in the above equation if $n$ is defined as the initial power law index of the entropy perturbations.

For flat cosmological constant($\Lambda$)-dominated models, however, the contribution of the ISW term is dominant on large scales for small $\Omega$, and this modifies the temperature spectrum [14]. For open models, in addition to the ISW effect, the cutoff in the $k$ integration and the modification of the Poisson equation near the curvature scale add further complications. For open adiabatic models, semi-analytic calculations on large scales have been recently performed by Kamionkowski and Spergel [7], who however only included the SW and ISW terms and therefore were unable to accurately probe the Tenerife scale. For BDM models, analytic formulae for $C_\ell$ were given by Gorski and Silk [15], who included the effects of geometry but neglected the contribution of the ISW effect. Moreover the favored range for the power law index for BDM models ($-1.5$ to $0$) from numerical simulations [16], is larger than the value expected by inflation ($-3$), and is out of the region of validity of the analytic formulation by the Gamma function. Gorski, Silk and Vittorio [17] examined large angular scale CMB anisotropies in cosmological constant–dominated adiabatic models, again including only the SW and ISW terms. There has been no previous work on large angular scale anisotropies in $\Lambda$–dominated BDM models.

The interplay of the ISW term, which enhances large–scale power, and geometry, which suppresses it via what in effect is gravitational focussing, is sufficiently intricate that we have been motivated to perform a full range of numerical calculations with both adiabatic and isocurvature initial conditions. We investigate the behaviour of $C_\ell$ with varying $\Omega$, and include the effects of varying the primordial index, the thermal history, and the vacuum contribution in spatially flat models. We use the gauge invariant method [12,18] to treat perturbation variables. The perturbation equations are solved numerically until the present epoch; detailed numerical treatments are given in Sugiyama and Gouda [19].

## III. Numerical Results

First, we present numerical results for CDM with adiabatic perturbations. Here we take $\Omega_B = 0.03$ and the dimensionless Hubble constant normalized by 100km/s/Mpc to $h = 0.5$. In figure 1, the $C_\ell$'s for different $\Omega$ are plotted (a) for open and (b) for flat $\Lambda$ models. Each $C_\ell$ distribution is normalized to the quadrupole anisotropy. The initial conditions for these models are taken as $|\delta\rho/\rho|^2 \propto \tilde{k}$, namely a Harrison–Zeldovich spectrum in an Einstein–de Sitter universe. For open universe models, this is not the only reasonable choice of initial conditions, the $C_\ell$'s having weak dependence on the different scaling of initial spectrum [7]. However we also consider an initial conditions produced by a low $\Omega$ inflationary model [20]. This initial spectrum is proportional to $\tilde{k}^{-1}$ on scales larger than the curvature scale. while it coincides with the Harrison-Zeldovich spectrum on small scales. The hatched region is the expected power law slope $1.4 \pm 0.6$ from the combined 1st and 2nd year COBE data [2,3]. The width of the window functions up to the half–power points are $l \leq 11$ and $13 \leq \ell \leq 30$ for the COBE and Tenerife experiments, respectively. Note that *all* $\Lambda$ models have effective $n$ smaller than unity on the COBE/Tenerife scales due to the ISW terms. Because of the curvature effect, $n_{\text{eff}} > 1$ for the open Harrison-Zeldovich models on the COBE/Tenerife scales. On the other hand, we get $n_{\text{eff}} < 1$ for the model of Ref. [20] because of the strong enhancement of large scale fluctuations. Even for this model, however, the curvature effect is apparent for the $\Omega = 0.1$ case.

The effect of reionization of the universe is shown in (c). Our adopted reionization model is the late–time fully ionized universe with electron–scattering optical depth unity. We compare reionized models with standard recombination models. Fluctuations on the COBE scale are not affected by reionization. However this is not the case on the Tenerife scale. There is a notable difference in particular for $\Lambda$ models. Because of the curvature effect on the geodesics, the horizon scale of the last scattering surface of an open model corresponds to a much smaller angle, i.e., larger $\ell$ than that of a $\Lambda$ model. For larger $\Omega$, the last scattering surface is earlier. This suggests that the $\Lambda$ model with small $\Omega$ has the greatest effect via reionization on large–scale CMB anisotropies for models with similar optical depths. In figure 1 (d), the dependence on the initial power law index $n$ is shown for the $\Omega = 0.1$ open and $\Lambda$ models, and the $\Omega = 1.0$ model. Even if $n = 1.5$, the $\Lambda$ model cannot be easily reconciled with the COBE slope. In order to more directly compare with observations, we show the expected temperature fluctuations for quadrupole, FIRS (Far Infra–Red Survey) [11] and Tenerife scales normalized to the COBE 10 degree scale in Table I. The effective power–law slopes $n_{\rm eff}$ on the COBE scale for each model are also shown. Using the relation between $C_\ell$ and $n$ for $\Omega = 1$ models, we define this $n_{\rm eff}$ by taking the ratio of $C_\ell$ at $\ell = 10$ and $\ell = 2$.

The $C_\ell$'s for BDM with primordial isocurvature perturbations are shown for different $\Omega$ in figure 2. Figures 2 (a) and (b) are fully ionized models with $n = -1$ for open and $\Lambda$ models, respectively. Here the initial power law index $n$ is defined as $|S|^2 \propto \tilde{k}^n$. In figure 2 (c), models with no reionization for $\Omega = 0.1$ open and $\Lambda$ models, and the $\Omega = 1.0$ model, are plotted together with the corresponding fully ionized models. In this figure, $n$ is also taken as $-1$. The dependence on $n$ is shown in (d) for fully ionized $\Omega = 0.1$ open and $\Lambda$ models and the $\Omega = 1.0$ model. It should be noticed that for both open and $\Lambda$ models, the dependence of $C_\ell$ on $n$ is weak. The thermal history of the universe is also weakly affected in open models as shown in (c). However the shape of the $C_\ell$ distribution for $\Lambda$ and $\Omega = 1.0$ models is very sensitive to the thermal history. In BDM models, the last scattering surface is much closer to the present for larger $\Omega$. Together with the geodesic effect, the open low $\Omega$ model is least affected by reionization on very large scales. The expected fluctuations of the quadrupole, FIRS and Tenerife anisotropies and the effective $n$ are shown in Table II.

## IV. Conclusions

In this *Letter*, we have investigated the shape of CMB anisotropies on very large scales for models with adiabatic and isocurvature initial conditions. On such large scales, cosmic variance cannot be negligible. Before summarizing our results, we discuss the effects of cosmic variance. We can assume that the $\ell$th moment of the expected temperature fluctuations obeys $\chi^2$ statistics with $2\ell + 1$ degrees of freedom. The 90% confidence region of the expected temperature fluctuations on scale $\ell$ is expressed in terms of the rms temperature fluctuations as $\alpha < \Delta T(\ell)/\Delta T_{rms}(\ell) < \beta$. Here $(\Delta T_{rms}(\ell)/T)^2 \equiv (2\ell + 1)C_\ell/4\pi$. $\alpha$ and $\beta$ are functions of $\ell$. If $\ell$ is large enough, $\alpha = \sqrt{1 - 1.96\sqrt{2/(2\ell + 1)}}$ and $\beta = \sqrt{1 + 1.96\sqrt{2/(2\ell + 1)}}$ because of the Gaussian nature of the $\chi^2$ distribution with many degrees of freedom. For small $\ell$, $(\ell, \alpha, \beta)$ =(2,0.48,1.49), (5,0.65,1.34), (10,0.74,1.25), (20, 0.82, 1.18). In tables I and II, we show the rms temperature fluctuations. Even though the quadrupole anisotropy contains a large cosmic variance, models with rms quadrupole anisotropy larger than $12\mu$K would be ruled out by the COBE detection, if this result is confirmed. Our normalization of fluctuations to the COBE 10 degree scale involves 30% cosmic variance. On the Tenerife scale, the effect of cosmic variance is less than 20%.

For adiabatic fluctuations, the difference between open models and $\Lambda$–dominated models is significant. Flat $\Lambda$–dominated models, which are favorable for large–scale structure formation, appear to be unable to account for the new COBE results and the Tenerife detection. However we caution that a precise comparison must be made using our non–power–law power spectrum before any definitive conclusions can be drawn. The inflationary open models [20] provide rather similar $C_\ell$'s to $\Lambda$–dominated models. These models will be discussed in detail in a forthcoming paper. Open Harrison-Zeldovich models are well fitted to COBE results, but have difficulty in producing large enough fluctuations on the Tenerife scale (Table I). The standard $\Omega = 1$ model has a slightly tilted slope $n_{\text{eff}} \simeq 1.1$ even on the COBE scale.

Isocurvature perturbations reveal intrinsically different shapes for CMB anisotropies on large scales. The value of $n_{\text{eff}} \simeq 2$ required by viable BDM models ($-1.5 \leq n \leq -0.5$) is only marginally consistent with the observational data. Open and $\Lambda$-dominated models have different effective $n$. For $\Lambda$-dominated models, the dependence on thermal history is important. Other parameter dependences, and in particular the initial $n$ dependences, are quite weak on the COBE scales for both open and $\Lambda$-dominated models. Geometry dominates over ISW in the absence of any intrinsic curvature fluctuations. BDM models may be distinguishable from models with adiabatic perturbations via the large–scale CMB anisotropies. The third and fourth–year COBE results and new large–scale experiments should provide a definitive probe of curvature in a BDM universe that may be written on the sky.

## Acknowledgements

The authors would like to thank W. Hu, M. Kamionkowski, D.H. Lyth, B. Ratra and M. White for valuable discussions. This research has been supported at Berkeley in part by grants from NASA and NSF. N.S. acknowledges financial support from a JSPS Postdoctoral Fellowship for Research Abroad.

TABLE I

| Expected $\Delta T(\mu K)$ of CDM adiabatic models | | | | |
|---|---|---|---|---|
| $\Omega$ | Q | FIRS | Tenerife | $n_{eff}$ |
| open | | | | |
| 0.1 | 12.1 | 39.4 | 25.3 | 1.5 |
| 0.1* | 11.6 | 39.8 | 26.4 | 1.6 |
| 0.1† | 15.4 | 35.5 | 18.1 | 0.84 |
| 0.2 | 13.8 | 37.7 | 22.1 | 1.1 |
| 0.2† | 16.7 | 35.3 | 17.7 | 0.63 |
| 0.3 | 14.6 | 37.7 | 22.3 | 1.0 |
| 0.3† | 16.8 | 36.0 | 19.2 | 0.68 |
| 0.4 | 14.8 | 38.5 | 23.7 | 1.0 |
| 0.6 | 13.2 | 40.8 | 27.5 | 1.4 |
| 0.8 | 11.4 | 41.3 | 28.3 | 1.6 |
| 1.0 | 14.7 | 38.8 | 24.3 | 1.1 |
| 1.0* | 14.8 | 37.3 | 21.8 | 1.0 |
| $\Lambda$ | | | | |
| 0.1 | 16.7 | 38.1 | 22.5 | 0.72 |
| 0.1* | 17.0 | 35.2 | 17.7 | 0.64 |
| 0.2 | 16.0 | 38.3 | 23.0 | 0.84 |
| 0.3 | 15.4 | 38.6 | 23.6 | 0.94 |
| 0.4 | 15.0 | 38.8 | 24.0 | 1.0 |
| 0.6 | 14.7 | 39.0 | 24.4 | 1.1 |
| 0.8 | 14.6 | 38.9 | 24.5 | 1.1 |
| Obs. | 6±3 | 45±13 | 42±9 | 1.4±0.6 |

* models with reionization ($\tau = 1$).
†models with initial power spectrum of Ref. [20].

TABLE II

Expected $\Delta T(\mu K)$ of BDM isocurvature models

| $\Omega$ | Q | | | | FIRS | | | | Tenerife | | | | $n_{\text{eff}}$ | | | |
|---|---|---|---|---|---|---|---|---|---|---|---|---|---|---|---|---|
| n | -1.5 | -1.0 | -0.5 | 0 | -1.5 | -1.0 | -0.5 | 0 | -1.5 | -1.0 | -0.5 | 0 | -1.5 | -1.0 | -0.5 | 0 |
| open | | | | | | | | | | | | | | | | |
| 0.1 | 10.2 | 9.6 | 9.1 | 8.6 | 44.7 | 47.8 | 50.7 | 53.4 | 33.6 | 37.4 | 40.5 | 43.1 | 1.9 | 2.0 | 2.1 | 2.1 |
| 0.1* | 10.1 | 9.7 | 9.1 | 8.3 | 49.8 | 52.8 | 55.6 | 58.8 | 38.9 | 41.7 | 44.3 | 47.2 | 1.9 | 1.9 | 2.0 | 2.2 |
| 0.2 | 10.3 | 9.6 | 9.1 | 8.8 | 43.4 | 47.2 | 50.5 | 53.2 | 31.7 | 36.4 | 40.0 | 42.6 | 1.8 | 2.0 | 2.1 | 2.1 |
| 0.3 | 10.5 | 9.5 | 9.0 | 8.8 | 42.8 | 47.1 | 50.9 | 53.8 | 30.5 | 35.9 | 40.1 | 42.8 | 1.8 | 2.0 | 2.1 | 2.1 |
| 0.4 | 10.7 | 9.5 | 8.9 | 8.7 | 42.4 | 47.2 | 51.7 | 55.0 | 29.7 | 35.7 | 40.5 | 43.6 | 1.7 | 2.0 | 2.1 | 2.1 |
| 0.6 | 11.2 | 9.5 | 8.6 | 8.3 | 42.0 | 47.6 | 53.9 | 59.2 | 28.5 | 35.5 | 42.0 | 46.7 | 1.6 | 2.0 | 2.2 | 2.2 |
| 0.8 | 11.7 | 9.8 | 8.3 | 7.4 | 41.6 | 48.0 | 56.2 | 65.1 | 27.6 | 35.5 | 43.7 | 51.3 | 1.5 | 1.9 | 2.2 | 2.4 |
| 1.0 | 12.3 | 10.2 | 8.4 | 6.9 | 41.2 | 48.1 | 57.6 | 69.8 | 26.8 | 35.2 | 44.6 | 54.8 | 1.4 | 1.8 | 2.2 | 2.5 |
| 1.0* | 7.8 | 6.2 | 5.3 | 5.0 | 55.9 | 64.6 | 72.1 | 77.2 | 46.1 | 54.3 | 60.5 | 64.1 | 2.3 | 2.7 | 2.9 | 2.9 |
| $\Lambda$ | | | | | | | | | | | | | | | | |
| 0.1 | 12.8 | 11.9 | 11.6 | 11.7 | 37.5 | 40.2 | 42.7 | 44.6 | 21.9 | 26.1 | 29.5 | 31.5 | 1.3 | 1.5 | 1.6 | 1.6 |
| 0.1* | 9.8 | 8.8 | 8.4 | 8.3 | 48.5 | 55.1 | 61.6 | 66.6 | 38.0 | 45.1 | 50.8 | 54.3 | 1.9 | 2.1 | 2.2 | 2.2 |
| 0.2 | 12.6 | 11.5 | 11.2 | 11.3 | 38.4 | 41.7 | 44.9 | 47.0 | 23.1 | 27.9 | 31.8 | 33.6 | 1.4 | 1.6 | 1.7 | 1.6 |
| 0.3 | 12.4 | 11.1 | 10.7 | 10.8 | 39.1 | 43.1 | 47.1 | 49.8 | 24.0 | 29.6 | 34.1 | 36.4 | 1.4 | 1.7 | 1.8 | 1.7 |
| 0.4 | 12.3 | 10.8 | 10.2 | 10.2 | 39.6 | 44.3 | 49.2 | 52.9 | 24.7 | 30.9 | 36.3 | 39.4 | 1.4 | 1.7 | 1.8 | 1.8 |
| 0.6 | 12.2 | 10.4 | 9.3 | 8.9 | 40.4 | 46.1 | 53.1 | 59.7 | 25.7 | 33.0 | 40.2 | 45.8 | 1.4 | 1.8 | 2.0 | 2.1 |
| 0.8 | 12.2 | 10.2 | 8.6 | 7.6 | 40.9 | 47.3 | 56.0 | 66.0 | 26.4 | 34.4 | 43.0 | 51.5 | 1.4 | 1.8 | 2.1 | 2.3 |
| Obs. | 6±3 | | | | 45±13 | | | | 42±9 | | | | 1.4±0.6 | | | |

* models with standard recombination (no reionization).

# Figure Captions

Figure 1: Power spectrum of temperature anisotropies $\ell(\ell+1)C_\ell$ normalized at $\ell = 2$ as a function of $\ell$ for adiabatic CDM. Open and $\Omega + \lambda = 1$ models ($\lambda \equiv \Lambda/3H_0^2$) with $\Omega = 0.1, 0.2, 0.3, 0.4, 0.6, 0.8$ and $1.0$ are shown in (a) and (b), respectively. Bold solid lines are $\Omega = 0.1$ and $\Omega = 1$. Dashed lines in (a) are low $\Omega$ models with initial power spectrum of Ref. [20]. $\Omega = 0.1$ open and $\Lambda$ models and $\Omega = 1$ model with optical depths $\tau = 1$ and $\tau = 0$ (no reionization) are shown in (c). Same models with initial power law index $n = 1$ and $n = 1.5$ are shown in (d). The hatched region is the expected power law slope $1.4 \pm 0.6$ from the combined 1st and 2nd year COBE data [2,3].

Figure 2: Same as figure 1 for isocurvature BDM. No recombination open and $\Omega + \lambda = 1$ models with $\Omega = 0.1, 0.2, 0.3, 0.4, 0.6, 0.8$ and $1.0$ are shown in (a) and (b), respectively. In (a), the $\Omega = 1$ adiabatic CDM model with $n = 1$ is also plotted for comparison. Bold solid lines are $\Omega = 0.1$ and $\Omega = 1$. $\Omega = 0.1$ open and $\Lambda$ models and $\Omega = 1$ model with no recombination and standard recombination (no reionization) are shown in (c). No recombination $\Omega = 0.1$ open and lambda models and $\Omega = 1$ model with initial power law index $n = -1.5, -1$ and $-0.5$ are shown in (d).

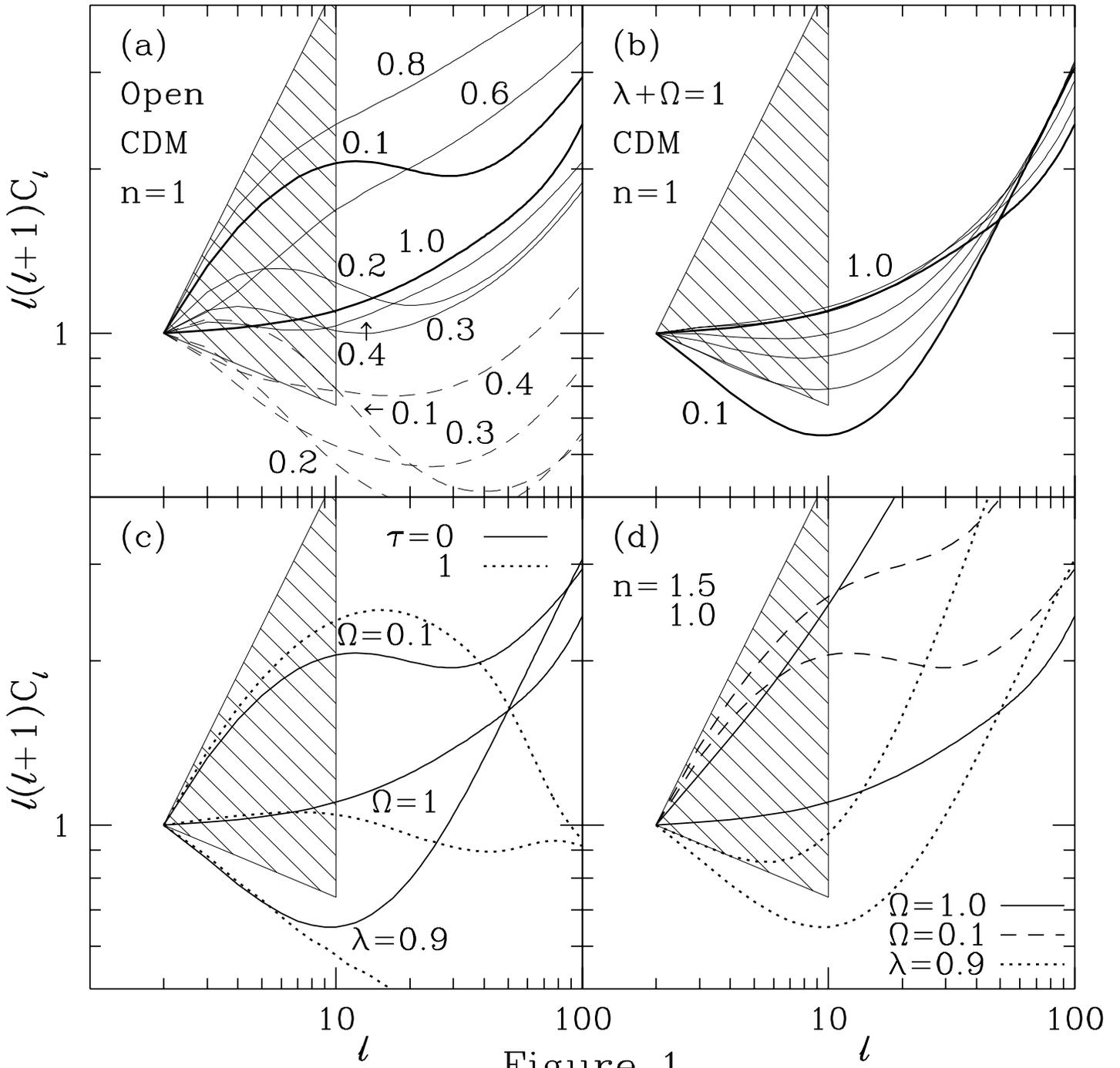

Figure 1

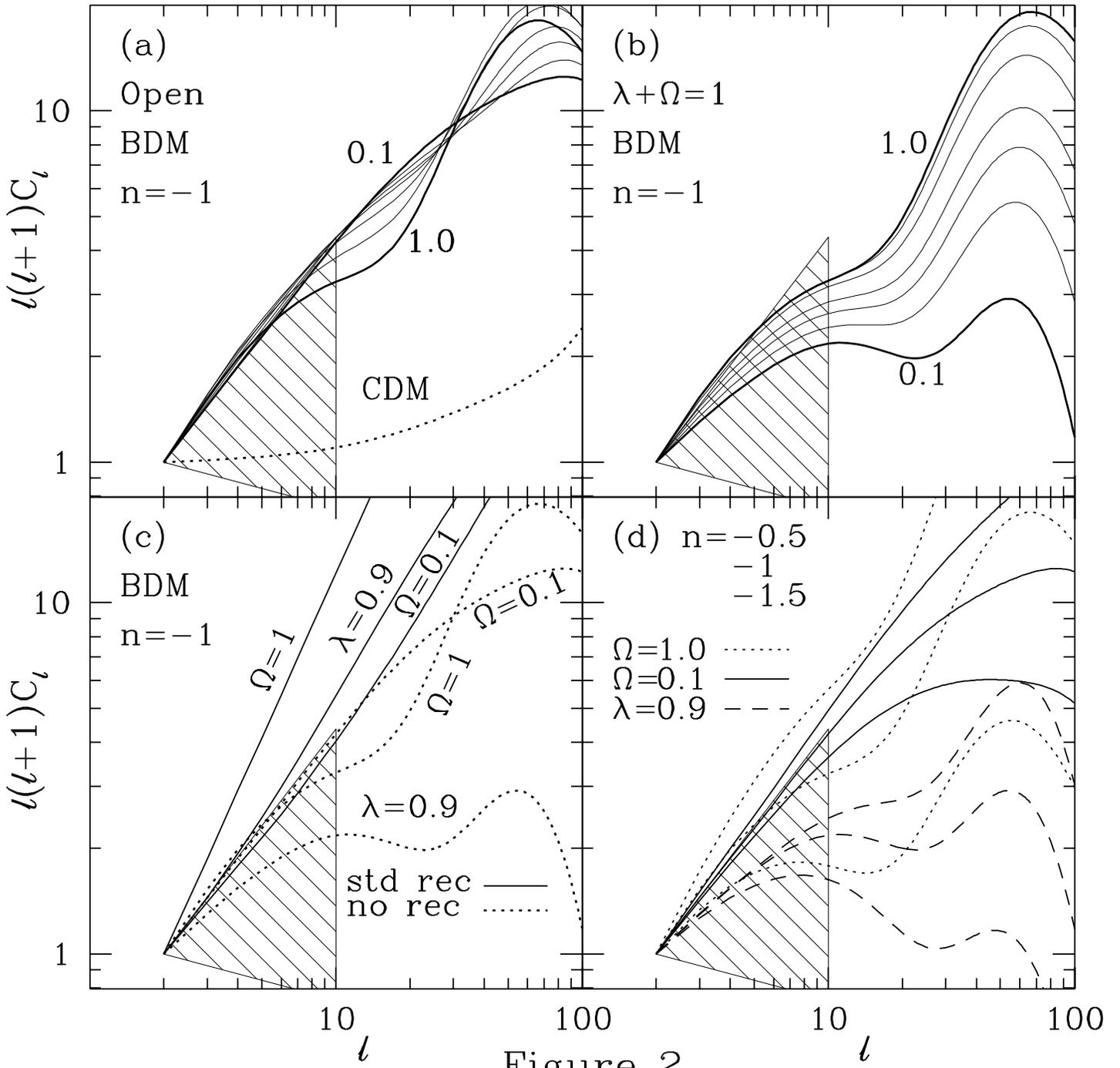

Figure 2